\documentclass[10pt]{article} 
\usepackage[preprint]{tmlr}

\usepackage{amsmath,amsfonts,bm}

\def\eqref#1{equation~\ref{#1}}

\def\1{\bm{1}}

\DeclareMathAlphabet{\mathsfit}{\encodingdefault}{\sfdefault}{m}{sl}
\SetMathAlphabet{\mathsfit}{bold}{\encodingdefault}{\sfdefault}{bx}{n}

\usepackage{hyperref}
\usepackage{url}
\usepackage[T1]{fontenc}
\usepackage{amsmath, amssymb, amsfonts, amsthm}
\usepackage{booktabs}
\usepackage{graphicx}
\usepackage{subcaption}

\usepackage{float}
 
\usepackage[utf8]{inputenc} 
\usepackage{hyperref}       
\usepackage{url}            
\usepackage{booktabs}       
\usepackage{amsfonts}       
\usepackage{nicefrac}       
\usepackage{microtype}      
\usepackage{xcolor}         
\usepackage{multirow}

\title{Can General-Purpose Omnimodels Compete with Specialists? A Case Study in Medical Image Segmentation}

\author{\name Yizhe Zhang \email zhangyizhe@njust.edu.cn \\
      \addr School of Computer Science and Engineering\\
      Nanjing University of Science and Technology
      \AND
      \name Qiang Chen \email chen2qiang@njust.edu.cn\\
      \addr School of Computer Science and Engineering\\
      Nanjing University of Science and Technology
      \AND
      \name Tao Zhou \email taozhou@njust.edu.cn\\
      \addr School of Computer Science and Engineering\\
      Nanjing University of Science and Technology}

\def\month{MM}  
\def\year{YYYY} 
\def\openreview{\url{https://openreview.net/forum?id=XXXX}} 

\begin{document}

\maketitle

\begin{abstract}
The emergence of powerful, general-purpose omnimodels capable of processing diverse data modalities has raised a critical question: can these ``jack-of-all-trades'' systems perform on par with highly specialized models in knowledge-intensive domains? This work investigates this question within the high-stakes field of medical image segmentation. We conduct a comparative study analyzing the zero-shot performance of a state-of-the-art omnimodel (Gemini, the ``Nano Banana'' model) against domain-specific deep learning models on three distinct tasks: polyp (endoscopy), retinal vessel (fundus), and breast tumor segmentation (ultrasound). Our study focuses on performance at the extremes by curating subsets of the ``easiest'' and ``hardest'' cases based on the specialist models' accuracy. Our findings reveal a nuanced and task-dependent landscape. For polyp and breast tumor segmentation, specialist models excel on easy samples, but the omnimodel demonstrates greater robustness on hard samples where specialists fail catastrophically. Conversely, for the fine-grained task of retinal vessel segmentation, the specialist model maintains superior performance across both easy and hard cases. Intriguingly, qualitative analysis suggests omnimodels may possess higher sensitivity, identifying subtle anatomical features missed by human annotators. Our results indicate that while current omnimodels are not yet a universal replacement for specialists, their unique strengths suggest a potential complementary role with specialist models, particularly in enhancing robustness on challenging edge cases.
\end{abstract}

\section{Introduction}
The field of artificial intelligence is undergoing a paradigm shift, moving from narrow models trained for specific tasks to large-scale, pre-trained foundation models. The latest evolution of this trend is the emergence of ``omnimodels,'' such as those in the Gemini family \citep{team2023gemini}, which are multimodal and can perform a vast array of tasks involving text, images, audio, and video with remarkable zero-shot and few-shot capabilities. Their impressive performance on general-purpose benchmarks has fueled speculation about their potential to disrupt specialized domains.

Among the most critical and demanding of these domains is medical image analysis. For decades, progress in this field has been driven by the development of highly specialized models, exemplified by architectures like the U-Net \citep{ronneberger2015u} and Swin-UNet\citep{cao2022swin}, which are designed and trained on curated datasets for a single, well-defined task, such as segmenting tumors in MRI scans or polyps in colonoscopy images. The success of these specialist models has been transformative, often achieving human-level performance and becoming integral to clinical research and diagnostic workflows.

This juxtaposition of paradigms raises a fundamental and timely research question: Can the broad, generalized knowledge of an omnimodel compete with the focused expertise of a specialist model in a domain where precision, reliability, and safety are paramount? While omnimodels promise versatility and reduced reliance on task-specific training data, it remains unclear whether they can match the nuanced and fine-grained perceptual capabilities required for complex medical segmentation tasks across different imaging modalities.

In this paper, we present three case studies involving polyp segmentation in endoscopic images, retinal vessel segmentation in fundus images, and breast tumor segmentation in ultrasound images, where we compare the zero-shot performance of Gemini with specialist models. For tasks involving relatively salient objects, even in noisy modalities like ultrasound, the omnimodel shows surprising robustness on hard cases that cause specialist models to fail completely. In contrast, for the intricate task of retinal vessel segmentation, the specialist model's architectural priors and focused training give it a decisive advantage across all levels of difficulty. Furthermore, our qualitative analysis uncovers instances where the omnimodel appears to identify fine anatomical structures missed in the ground-truth annotations, suggesting a potential for higher sensitivity. These results provide a nuanced perspective on the current capabilities of omnimodels in medicine, highlighting both their potential and their present limitations.

\section{Background and Related Work}

Our work is positioned at the intersection of three active research areas: the development of specialist models for medical segmentation, the emergence of general-purpose foundation models, and the critical analysis of model robustness.

\subsection{Specialist Models for Medical Image Segmentation}
For years, the state of the art in medical image segmentation has been advanced by highly specialized models. The U-Net architecture~\citep{ronneberger2015u} stands as a cornerstone in this domain, with its symmetric encoder-decoder structure and skip connections proving exceptionally effective for capturing both contextual information and fine-grained details from a limited number of training samples. This foundational design has inspired a vast lineage of variants tailored for specific modalities and anatomical structures. For the tasks investigated in our study, dedicated models have been developed to tackle their unique challenges. In polyp segmentation, for instance, models like HSNet~\citep{zhang2022hsnet} have emerged, which employ hybrid architectures combining Convolutional Neural Networks (CNNs) for local feature extraction with Transformers for capturing long-range dependencies, aiming to improve accuracy on polyps of varying shapes and sizes. Similarly, retinal vessel segmentation, a task requiring exquisite detail preservation, has seen the development of numerous specialized deep learning approaches designed to trace fine, low-contrast vascular structures~\citep{jin2022fives}. These specialist models represent the incumbent paradigm, characterized by focused training and architectural priors optimized for a single task.

\subsection{The Rise of Generalist Foundation Models}
A recent paradigm shift in artificial intelligence has been driven by the development of large-scale, pre-trained foundation models. In computer vision, the Segment Anything Model (SAM)~\citep{kirillov2023segment} marked a significant milestone, demonstrating remarkable zero-shot segmentation capabilities on natural images. This success immediately catalyzed a wave of research adapting SAM and its principles to the medical domain, resulting in models like MedSAM and numerous other variants~\citep{ma2023segment}. Concurrently, the evolution of Large Multimodal Models (LMMs), or ``omnimodels'' such as those in the Gemini family~\citep{team2023gemini}, Med-PaLM M~\citep{tu2024medpalm}, and anticipated next-generation systems like GPT-5~\citep{hu2025benchmarking}, has introduced the ability to perform complex reasoning and generation tasks across text, image, and other modalities. Critically, their advanced image generation capability is highly relevant to medical image segmentation, as it enables these models to go beyond classification and instead directly generate a segmentation mask as an image output. These models, often trained on vast, web-scale datasets, possess a broad world knowledge that allows them to attempt specialized tasks like medical image analysis in a zero-shot setting. Other related efforts, such as LLaVA-Med~\citep{li2023llava}, have focused on creating biomedical conversational assistants by fine-tuning general-domain vision-language models on curated medical data. This trend towards generalist systems raises the central question of our paper: to what extent can their broad, generalized capabilities rival the deep, narrow expertise of specialist models?

\subsection{Comparative Studies and Robustness Evaluation}
The burgeoning presence of generalist models has naturally led to comparative studies against their specialist counterparts. Recent comprehensive surveys have begun to map this new landscape, concluding that the performance trade-offs are complex and highly dependent on the specific task, modality, and dataset~\citep{moglia2025generalist}. Initial evaluations of LMMs like GPT-4 Vision on diagnostic tasks have revealed promising, yet inconsistent, zero-shot performance, often excelling at high-level interpretation but struggling with the fine-grained detail required for clinical-grade responses~\citep{wu2023can}. Further supporting this, a broad evaluation by Zhang et al.~\citep{zhang2024potential} across 14 medical imaging datasets found that while MLLMs like the GPT and Gemini series show potential, their performance is task-dependent; for instance, the GPT series was more proficient at lesion segmentation, whereas the Gemini series excelled in report generation and lesion detection.

In this study, we focus on medical image segmentation, specifically examining the omnimodel’s ability to generate segmentation outputs by taking a raw image as input and a text prompt specifying the target structure. In addition, most of previous comparisons rely on aggregate performance metrics across entire test sets. Such an approach can obscure critical differences in model behavior, particularly in failure modes. Our work diverges from this standard by focusing on performance at the extremes. The concept of analyzing model performance based on sample difficulty has been explored in the broader machine learning community, where it has been shown that deep networks often learn ``easy'' examples before mastering ``hard'' ones~\citep{mayo2023hard}. This perspective is crucial for understanding model robustness. Our study is also conceptually linked to the field of failure detection in medical imaging, which seeks to identify unreliable segmentations to ensure clinical safety~\citep{zenk2024comparative}. Instead of predicting failure post-hoc, our study provides a direct, comparative analysis of how generalist and specialist models behave on the very samples that are intrinsically easiest or hardest for the specialist, thereby offering a deeper insight into their respective robustness and failure characteristics. 
\section{Experimental Design}

To provide a focused comparison between specialist and generalist models, we designed a series of case studies centered on three distinct medical segmentation tasks. Our study is characterized by a performance stratification approach to probe model behavior at its operational extremes.

\textbf{Tasks and Datasets.} We selected three tasks with differing characteristics: 1) Polyp segmentation from endoscopic images using the CVC-ColonDB dataset~\citep{tajbakhsh2015automated}, which involves identifying relatively large, salient objects. 2) Retinal vessel segmentation from fundus images using the FIVE dataset~\citep{jin2022fives}, a task that requires delineating fine, intricate, low-contrast structures. 3) Breast tumor segmentation from ultrasound images using the Breast Ultrasound Images (BUSI) dataset~\citep{vallez2025bus_uclm}, a task that requires identifying lesions in a noisy and challenging modality.

\textbf{Model Selection.} For each task, we compared a representative domain-specialized network with a state-of-the-art general-purpose omnimodel.
\begin{itemize}
    \item \textbf{Specialist Models:} For polyp segmentation, we employed the pre-trained HSNet~\citep{zhang2022hsnet} provided by the original authors, specifically designed for endoscopic polyp analysis. For retinal vessel segmentation, we trained a standard U-Net architecture~\citep{ronneberger2015u} on the official training split of the FIVE dataset. For breast tumor segmentation, we utilized an in-house implementation of Mask2Former~\citep{cheng2022masked}, trained on samples from the BUSI ultrasound dataset.
    \item \textbf{Omnimodel:} As a generalist state-of-the-art model, we adopted Gemini, using its ``Create images'' tool (using the ``Nano Banana'' model) to assess zero-shot segmentation performance without any task-specific fine-tuning.
\end{itemize}

\textbf{Performance Stratification.} A core component of our study is the analysis of performance on the easiest and hardest cases. For each task, we first evaluated the corresponding specialist model on its entire test set. Based on the Dice similarity coefficient for each sample, we curated two subsets for our comparative analysis: the \textbf{top 5\% best-performing samples (``easy samples'')} and the \textbf{bottom 5\% worst-performing samples (``hard samples'')}. This allows for a direct comparison of how the omnimodel fares on cases where the specialist either excels or fails catastrophically.

\textbf{Prompting Strategy.} We interacted with Gemini via its image generation function using task-specific prompts designed to elicit a binary segmentation mask.
\begin{itemize}
    \item For polyp segmentation: ``Generate a binary segmentation mask of the polyp, ensuring the entire polyp region is fully captured without missing any parts.''
    \item For retinal vessel segmentation: ``Generate a binary segmentation mask of the blood vessels, ensuring the vessels are solid (no hollow or broken interiors).''
    \item For breast tumor segmentation: ``Generate a binary segmentation mask of breast ultrasound images, ensuring precise delineation of the lesion region.''
\end{itemize}
The output of the omnimodel was binarized using a simple threshold and subsequently compared with the ground truth masks for quantitative evaluation.\footnote{There is considerable potential to further optimize the prompt design to enhance performance, and we leave such exploration for future work.}

\textbf{Evaluation Metrics.} To quantitatively assess segmentation performance, we employed three widely used metrics: the Dice similarity coefficient (Dice) and the 95\% Hausdorff Distance (HD95). Dice measures the degree of overlap between the predicted and ground-truth masks, with a particular emphasis on the accurate delineation of the \textit{foreground} region of interest. HD95 evaluates boundary similarity by computing the 95th percentile of distances between predicted and ground-truth contours, which is especially relevant in medical imaging where precise boundary localization has clinical importance. Higher Dice score, and a lower HD95 value, indicate better performance.

\section{Case Studies and Results}

\subsection{Polyp Segmentation in Endoscopic Images}

Our first case study evaluated performance on the CVC-ColonDB dataset. Table~\ref{tab:segmentation_results} presents the quantitative comparison between the specialist HSNet and the generalist Gemini on the stratified easy and hard sample sets.

\begin{figure}[t]
    \centering
    \includegraphics[width=0.8\linewidth]{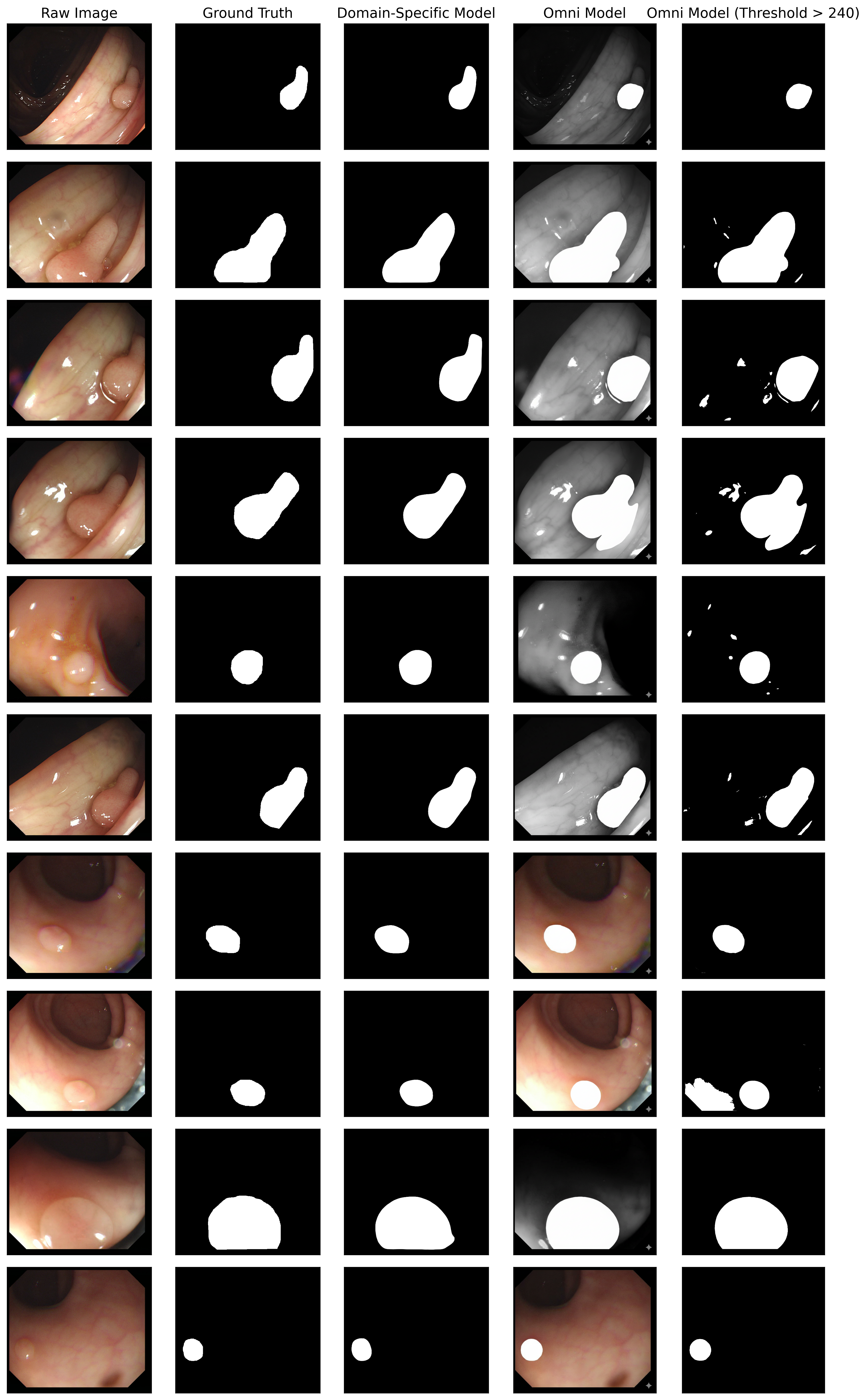}
    \caption{Segmentation Results for Easy Samples in CVC-ColonDB dataset.}
    \label{fig:colon-db-best}
\end{figure}

On the \textbf{easy samples}, the specialist HSNet model demonstrates superior performance, as expected. It achieves near-perfect scores with a Dice of 97.4\% and its extremely low HD95 (6.9 pixels) confirms its high boundary precision. While Gemini Nano Banana's performance is lower, its Dice score of 87.7\% is still commendable for a zero-shot task, showcasing its strong general-purpose capabilities. Visualizations in Figure~\ref{fig:colon-db-best} corroborate these findings, showing HSNet's masks closely aligning with the ground truth.

\begin{table}[t]
\centering
\caption{Performance comparison on the best and worst 5\% performing samples from the CVC-ColonDB test set. Dice is reported as percentages ($\uparrow$), while HD95 is in pixels ($\downarrow$).}
\label{tab:segmentation_results}
\begin{tabular}{@{}llcc@{}}
\toprule
\textbf{Sample Set} & \textbf{Model} & \textbf{Dice (\%) $\uparrow$} & \textbf{HD95 (pixels) $\downarrow$} \\
\midrule
\addlinespace[0.3em]
\textbf{Easy Samples} & Specialist (HSNet) & 97.4  & 6.9 \\
& Omnimodel (Gemini Nano Banana) & 87.7 & 87.4 \\
\addlinespace[0.6em]
\textbf{Hard Samples} & Specialist (HSNet) & 4.3  & 332.4 \\
& Omnimodel (Gemini Nano Banana) & 23.6 & 304.6 \\
\addlinespace[0.3em]
\bottomrule
\end{tabular}
\end{table}

\begin{figure}[t]
    \centering
    \includegraphics[width=0.8\linewidth]{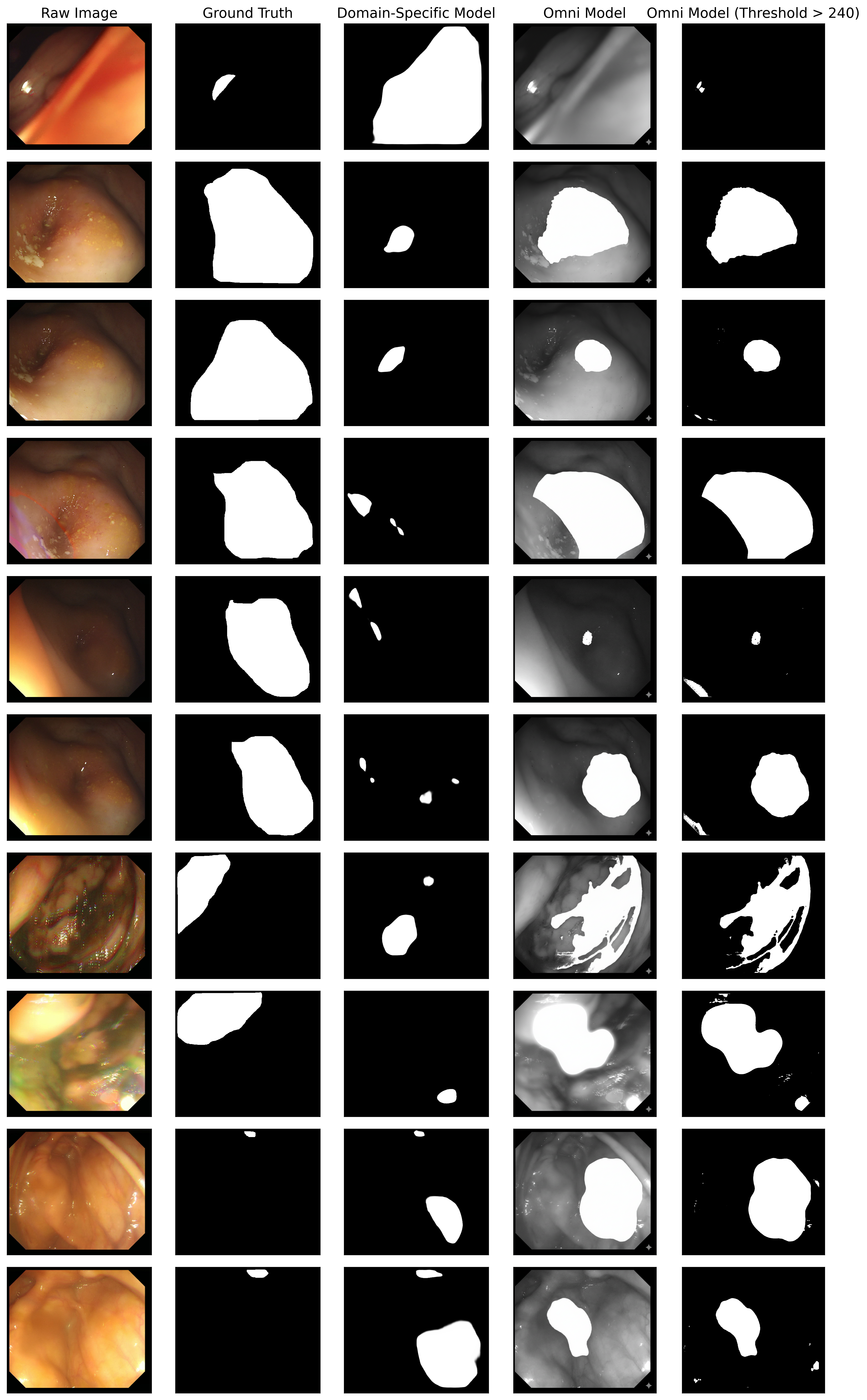}
    \caption{Segmentation Results for Hard Samples in CVC-ColonDB dataset.}
    \label{fig:colon-db-worst}
\end{figure}

The performance dynamic reverses on the \textbf{hard samples}. HSNet's performance collapses, with its Dice score plummeting to a mere 4.3\%, indicating a complete failure to segment the polyp, as visualized in Figure~\ref{fig:colon-db-worst}. In contrast, Gemini Nano Banana achieves a Dice score of 23.6\%. While modest, this result is substantially better than HSNet's and suggests that the omnimodel possesses greater robustness and generalization when faced with ambiguous or out-of-distribution examples that cause the specialist to fail. This highlights a potential advantage of omnimodels in handling challenging edge cases.

\subsection{Retinal Vessel Segmentation in Fundus Images}

For our second case study, we shifted to the fine-grained task of retinal vessel segmentation using the FIVE dataset. The results, summarized in Table~\ref{tab:retinal_results}, reveal a different relationship between the specialist and generalist models.

On both \textbf{easy and hard samples}, the specialized U-Net model maintains a decisive performance advantage. For easy cases, the U-Net achieves a Dice score of 91.5\%, significantly outperforming Gemini Nano Banana's 62.4\%. Unlike the polyp task, this performance gap persists on the hard samples. The U-Net achieves a Dice score of 58.7\% on these challenging images, more than double Gemini Nano Banana's score of 28.3\%.

\begin{table}[t]
\centering
\caption{Performance comparison on the best and worst 5\% performing samples from the FIVE test set. Dice is reported as percentages ($\uparrow$), while HD95 is in pixels ($\downarrow$).}
\label{tab:retinal_results}
\begin{tabular}{@{}llcc@{}}
\toprule
\textbf{Sample Set} & \textbf{Model} & \textbf{Dice (\%) $\uparrow$} & \textbf{HD95 (pixels) $\downarrow$} \\
\midrule
\addlinespace[0.3em]
\textbf{Easy Samples} & Specialist (U-Net) & 91.5  & 12.0 \\
& Omnimodel (Gemini Nano Banana) & 62.4  & 111.0 \\
\addlinespace[0.6em]
\textbf{Hard Samples} & Specialist (U-Net) & 58.7  & 174.3 \\
& Omnimodel (Gemini Nano Banana) & 28.3  & 408.9 \\
\addlinespace[0.3em]
\bottomrule
\end{tabular}
\end{table}

\begin{figure}[t]
    \centering
    \includegraphics[width=0.8\linewidth]{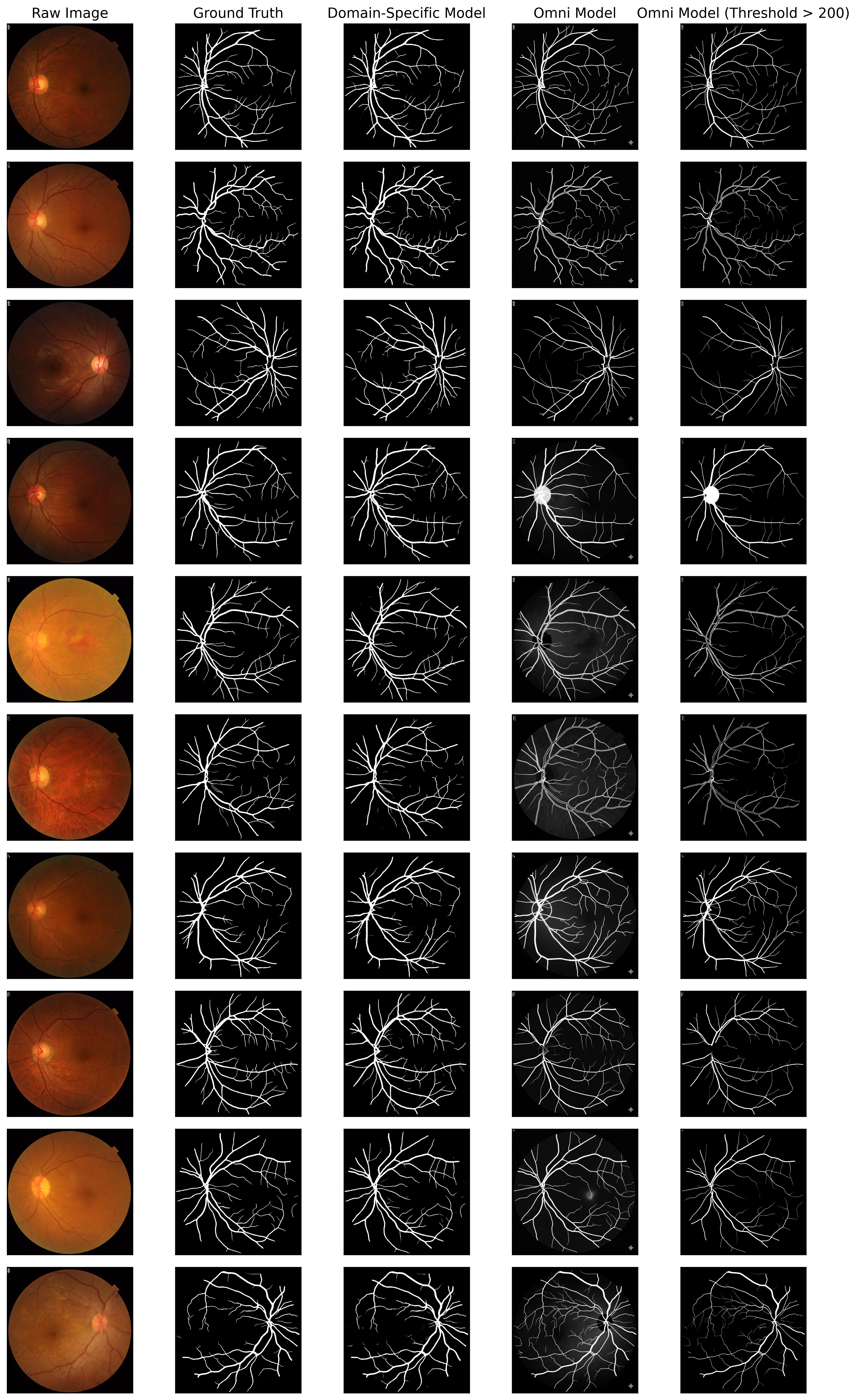}
    \caption{Segmentation Results for Easy Samples in FIVE dataset.}
    \label{fig:retinal_best}
\end{figure}

\begin{figure}[t]
    \centering
    \includegraphics[width=0.8\linewidth]{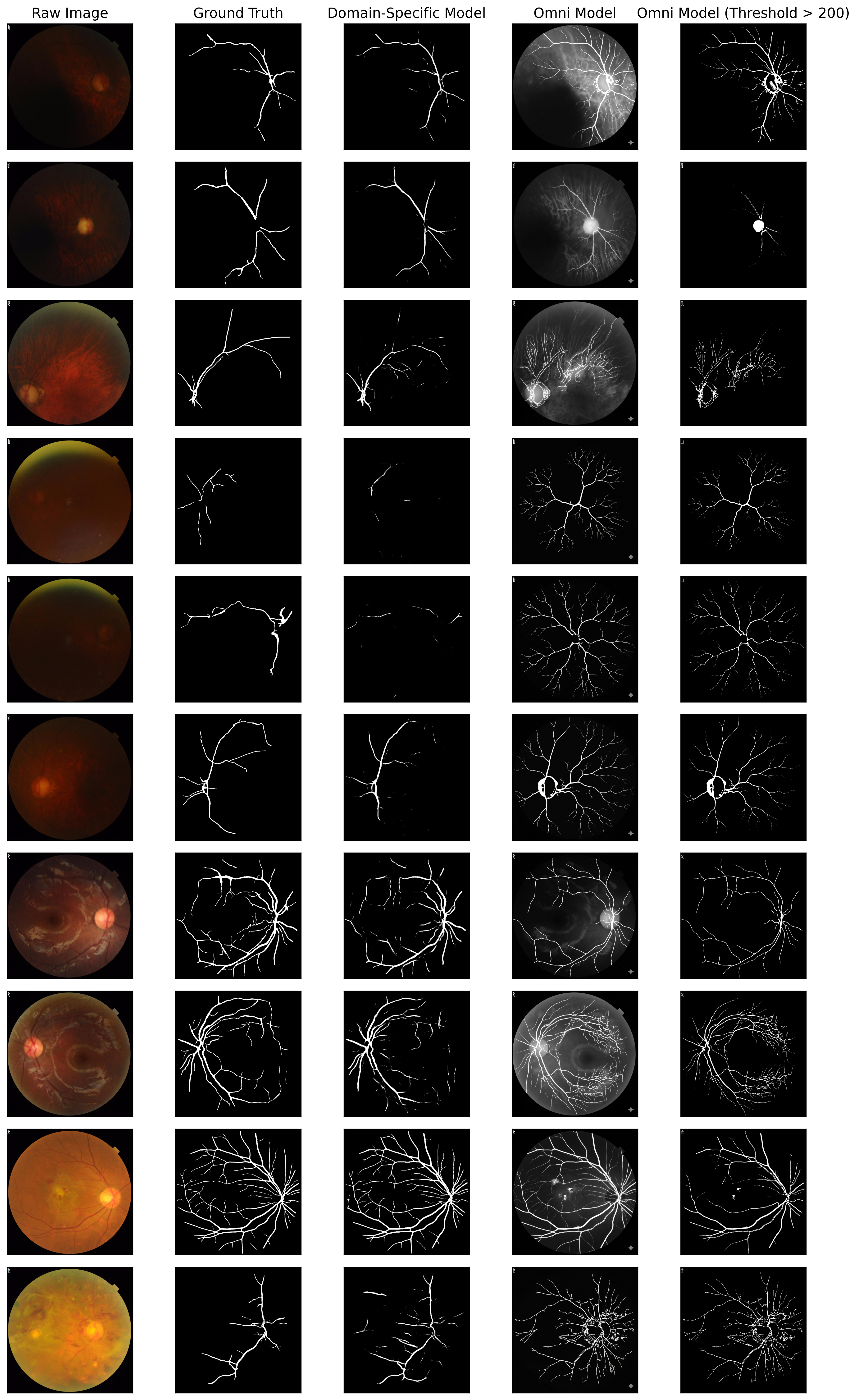}
    \caption{Segmentation Results for Hard Samples in FIVE dataset.}
    \label{fig:retinal_worst}
\end{figure}

Qualitative analysis from Figures~\ref{fig:retinal_best} and \ref{fig:retinal_worst} reinforces these quantitative findings. The U-Net produces clean, precise segmentations that capture fine vessel details, even in challenging, low-contrast images. The omnimodel, while localizing the main vascular structures, struggles with precision and detail. This suggests that for tasks requiring fine-grained perception of intricate, low-contrast patterns, the architectural priors and focused training of a specialist model are crucial for both accuracy and robustness.

\begin{figure}[t]
    \centering
    \includegraphics[width=0.5\linewidth]{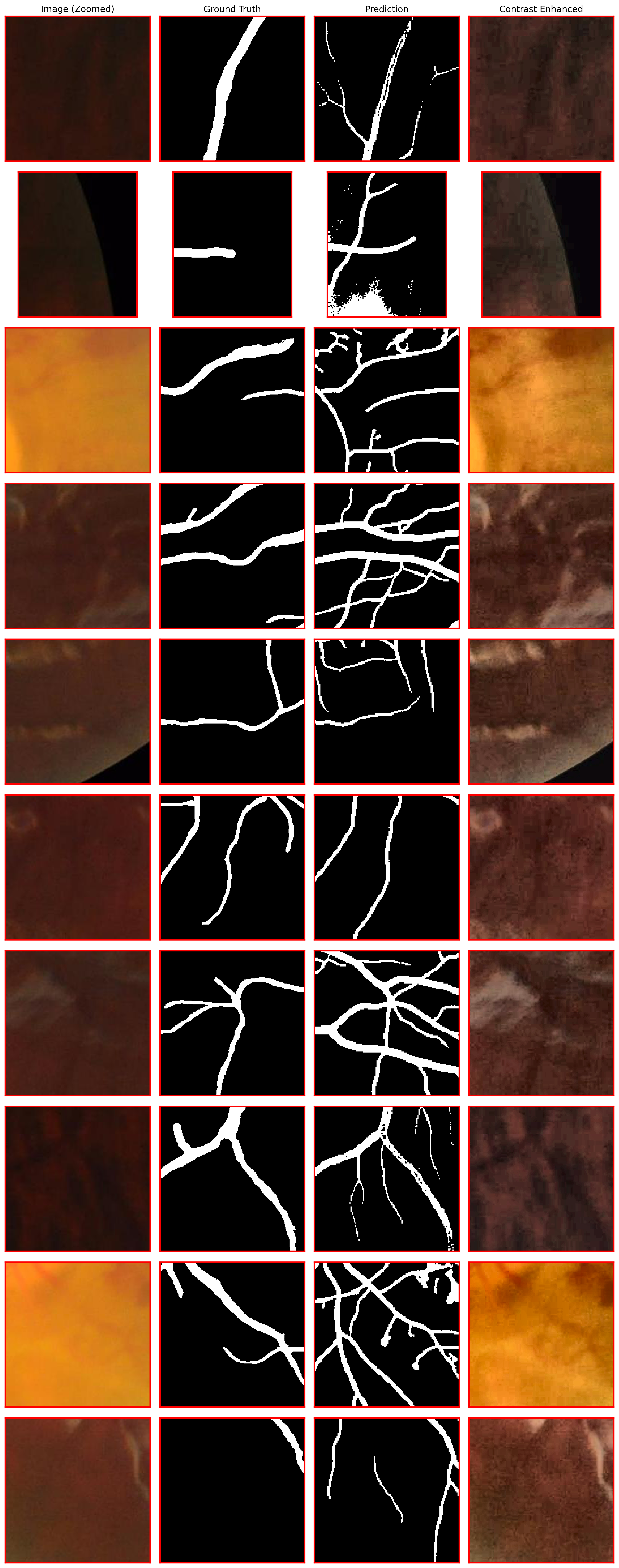}
    \caption{Zoomed-in qualitative comparison of retinal vessel segmentation. Gemini Nano Banana often identifies fine vessels (third column) missed in the ground truth (second column) but visible with contrast enhancement (fourth column).}
    \label{fig:retinal_zoom}
\end{figure}

Despite its lower quantitative scores, a notable qualitative finding emerged. As shown in Figure~\ref{fig:retinal_zoom}, Gemini Nano Banana often identifies fine vessels that are absent in the ground truth but become visible in contrast-enhanced versions of the original images. This suggests the omnimodel may possess a higher sensitivity for detecting minute details overlooked during manual annotation, a potentially valuable characteristic.

\subsection{Breast Tumor Segmentation in Ultrasound Images}
Our third case study addressed breast tumor segmentation on the BUSI dataset, a task characterized by low tissue contrast and imaging artifacts, demanding a high degree of expert interpretation. The results, shown in Table~\ref{tab:ultrasound_results}, reveal a performance pattern similar to that observed in polyp segmentation.

On \textbf{easy samples}, the specialist Mask2Former model demonstrates clear superiority, achieving a high Dice score of 95.2\% compared to Gemini Nano Banana’s 70.5\%. This confirms the specialist's effectiveness on straightforward, in-distribution examples.

However, on the \textbf{hard samples}, we again observe a role reversal. Mask2Former's performance catastrophically fails, dropping to a Dice score of just 3.1\%. This indicates its inability to handle challenging cases with ambiguous boundaries or unusual presentations. In contrast, Gemini Nano Banana shows better resilience, maintaining a Dice score of 21.1\%. Although this is not a high score, it reinforces the observation that the omnimodel's broad knowledge base provides a degree of robustness against the types of edge cases that cause narrowly trained models to fail. This result from a challenging, noisy modality further strengthens the hypothesis that omnimodels may have a unique advantage in handling difficult, out-of-distribution samples.

\begin{table}[t]
\centering
\caption{Performance comparison on the best and worst 5\% performing ultrasound segmentation samples. Dice is reported as percentages ($\uparrow$), while HD95 is in pixels ($\downarrow$).}
\label{tab:ultrasound_results}
\begin{tabular}{@{}llcc@{}}
\toprule
\textbf{Sample Set} & \textbf{Model} & \textbf{Dice (\%) $\uparrow$} & \textbf{HD95 (pixels) $\downarrow$} \\
\midrule
\addlinespace[0.3em]
\textbf{Easy Samples} & Specialist (Mask2Former) & 95.2 & 15.5 \\
& Omnimodel (Gemini Nano Banana) & 70.5  & 121.3 \\
\addlinespace[0.6em]
\textbf{Hard Samples} & Specialist (Mask2Former) & 3.1 & 263.8 \\
& Omnimodel (Gemini Nano Banana) & 21.1  & 198.1 \\
\addlinespace[0.3em]
\bottomrule
\end{tabular}
\end{table}

\begin{figure}[t]
    \centering
    \includegraphics[width=0.8\linewidth]{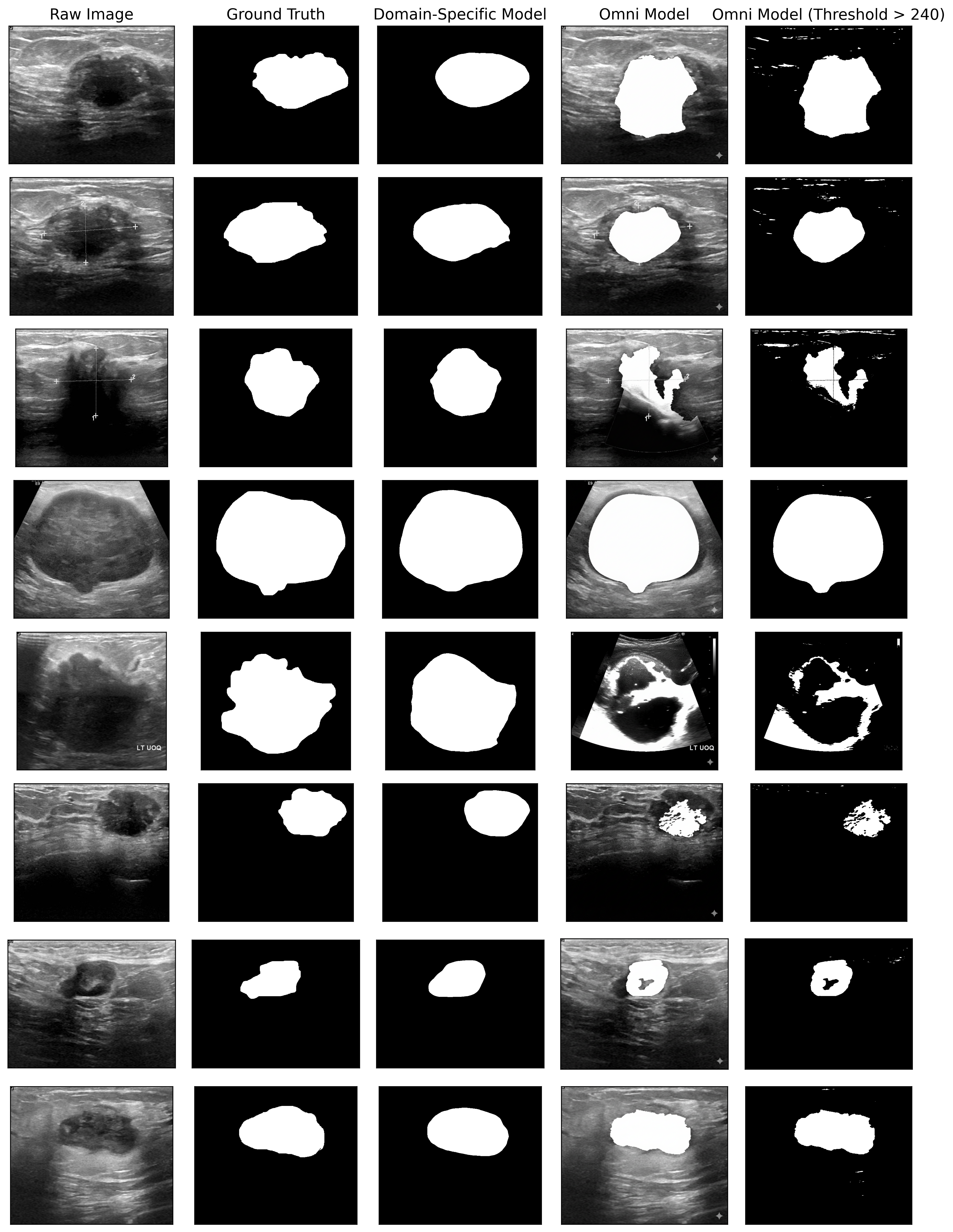}
    \caption{Segmentation Results for Easy Samples in BUSI dataset.}
    \label{fig:BUSI_best}
\end{figure}

\begin{figure}[t]
    \centering
    \includegraphics[width=0.8\linewidth]{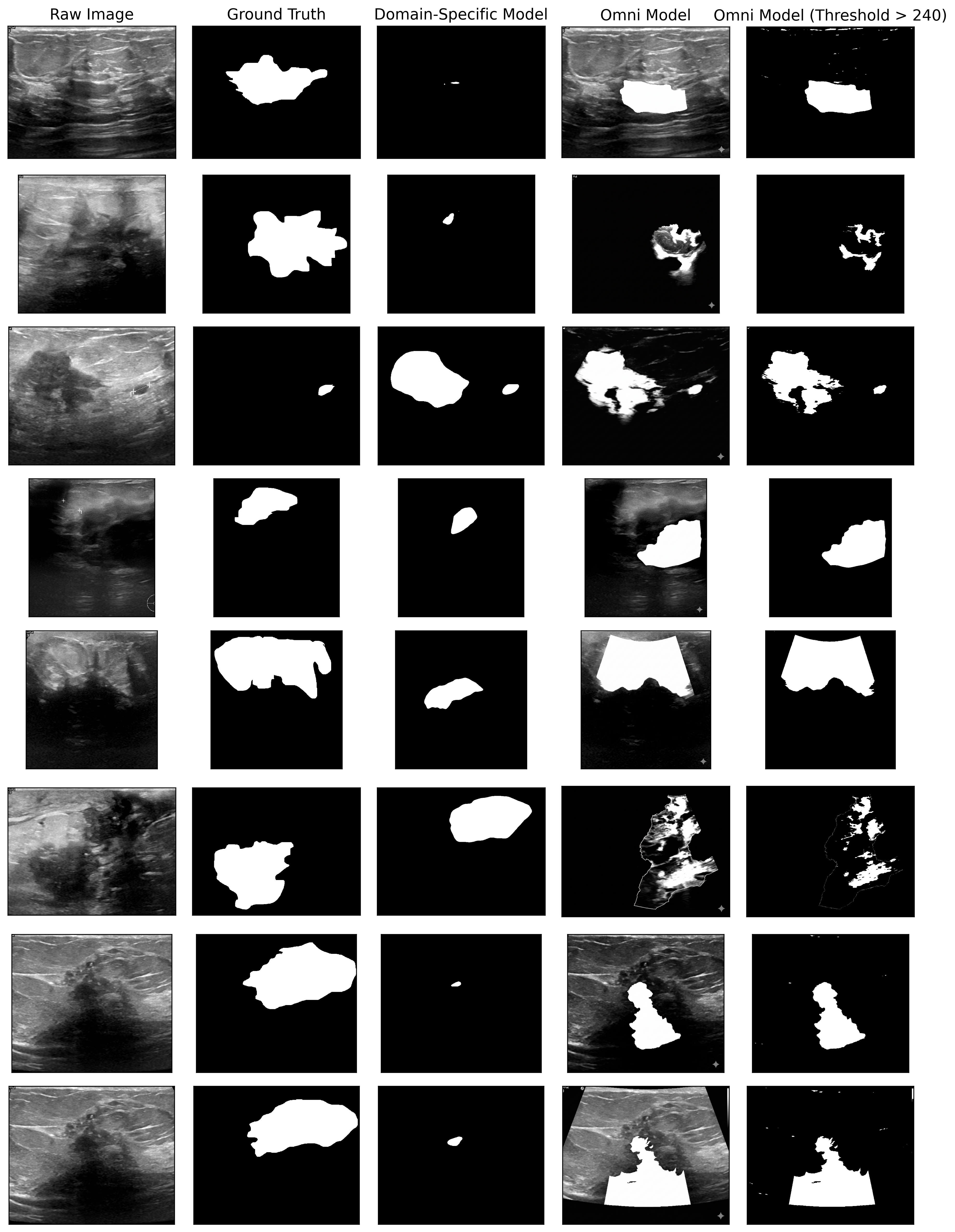}
    \caption{Segmentation Results for Hard Samples in BUSI dataset.}
    \label{fig:BUSI_worst}
\end{figure}

The visual results in Figures \ref{fig:BUSI_best} and \ref{fig:BUSI_worst} provide an illustration of these quantitative findings. For the easy samples (Figure \ref{fig:BUSI_best}), the domain-specific model's predictions are nearly indistinguishable from the ground truth, capturing the lesion boundaries with high fidelity. The omnimodel, while correctly identifying the lesion's general location, produces a much rougher and less precise object boundaries. For the hard samples (Figure \ref{fig:BUSI_worst}), the domain-specific model fails entirely, often producing a tiny, misplaced segmentation or nothing at all. In contrast, the omnimodel, despite the challenging image quality, generates a segmentation that, while still not good enough, is a far more reasonable approximation of the ground truth, demonstrating its superior robustness in these difficult scenarios.

\section{Discussion}

Our findings present a nuanced and task-dependent answer to the central question of whether omnimodels can compete with specialists in medical image segmentation. The results suggest that the current relationship is not one of direct competition, but rather one of complementary strengths and weaknesses, contingent on both the nature of the segmentation task and the difficulty of the specific case.

The consistent superiority of specialist models on ``easy'' samples across all three tasks is unsurprising. These models, through their specialized architectures and focused training, have been optimized to excel on the core distribution of their target data. They represent a paradigm of deep, narrow intelligence, and on their home turf, they remain the undisputed experts. This confirms that for routine, in-distribution tasks, purpose-built models are highly effective and reliable.

The most compelling insight emerges from the analysis of ``hard'' samples. For tasks involving the segmentation of relatively salient objects, such as polyps and breast tumors, the omnimodel demonstrated a striking degree of robustness where the specialist models failed catastrophically. We hypothesize that this resilience stems from the omnimodel's vast, generalized pre-training. While a specialist model learns specific priors about what a ``polyp'' or ``tumor'' looks like within its training set, an omnimodel has learned much broader, more fundamental concepts of objectness, boundaries, and textures from web-scale data. When faced with an atypical, out-of-distribution case that violates the specialist's narrow priors, the specialist's performance collapses. The omnimodel, in contrast, can fall back on its more generalized ``common sense'' visual understanding to produce a plausible, albeit imperfect, segmentation. This suggests a potential role for omnimodels as a safety net or a second-opinion system to handle the long tail of challenging cases that often confound specialized systems.

Conversely, the retinal vessel segmentation task highlights the current limitations of this generalized knowledge. Here, the task is not merely to identify an ``object'' but to trace fine, intricate, low-contrast linear patterns. The architectural priors of the U-Net, specifically designed for such multi-scale feature extraction, prove indispensable. The specialist model's superior performance on both easy and hard cases indicates that for tasks demanding high-fidelity, fine-grained perception, specialized designs and training remain paramount. The omnimodel's broader knowledge does not, in its current form, translate to the level of precision required for this domain.

Perhaps one of the most intriguing qualitative findings is the omnimodel's potential for higher sensitivity, as observed in the retinal vessel task where it identified structures missed in the ground-truth annotations. While this requires more rigorous validation, it opens up a fascinating possibility: omnimodels could potentially serve not just as segmentation tools, but as aids in the data curation process itself. They might be used to flag potential inconsistencies or missed annotations in existing datasets, thereby helping to create higher-quality ground truth for training the next generation of more accurate specialist models.

Finally, it is important to acknowledge the limitations of this study. Our analysis of the omnimodel was conducted in a purely zero-shot setting. Its performance could potentially be improved significantly with few-shot prompting or task-specific fine-tuning. Furthermore, our study utilized a single omnimodel, and results may vary with other emerging architectures. The definition of ``hard'' cases was also contingent on the failure of one specific specialist model for each task. Nevertheless, this ``performance-stratified'' approach provides a valuable lens through which to understand the distinct operational characteristics and failure modes of these two classes of models.

\section{Conclusion}

In this work, we examined whether a general-purpose omnimodel can rival specialized models in the demanding domain of medical image segmentation. Our case studies demonstrate that the answer is nuanced rather than a simple yes or no. Specialist models remain indispensable, excelling in in-distribution settings and tasks that demand fine-grained detail.

At the same time, we find that omnimodels exhibit greater robustness on difficult, out-of-distribution samples, particularly for salient object segmentation. Where specialist models may fail entirely, the omnimodel’s broad prior knowledge enables it to maintain a resilient baseline. This suggests not a future defined by competition, but by synergy. We envision hybrid systems in which specialist models deliver high accuracy and efficiency on routine cases, while omnimodels serve as a fallback mechanism or a ``second reader'' for challenging edge cases. Such collaboration can lead to safer and more reliable clinical AI systems. Moreover, the potential of omnimodels to help refine manually labeled datasets further underscores their value. Ultimately, the era of foundation models in medicine is not about replacing experts, but about equipping them with a more versatile and robust toolkit.

\bibliography{tmlr}

\begin{thebibliography}{18}
\providecommand{\natexlab}[1]{#1}
\providecommand{\url}[1]{\texttt{#1}}
\expandafter\ifx\csname urlstyle\endcsname\relax
  \providecommand{\doi}[1]{doi: #1}\else
  \providecommand{\doi}{doi: \begingroup \urlstyle{rm}\Url}\fi

\bibitem[Cao et~al.(2022)Cao, Wang, Chen, Jiang, Zhang, Tian, and Wang]{cao2022swin}
Hu~Cao, Yueyue Wang, Joy Chen, Dongsheng Jiang, Xiaopeng Zhang, Qi~Tian, and Manning Wang.
\newblock Swin-unet: Unet-like pure transformer for medical image segmentation.
\newblock In \emph{European conference on computer vision}, pp.\  205--218. Springer, 2022.

\bibitem[Cheng et~al.(2022)Cheng, Misra, Schwing, Kirillov, and Girdhar]{cheng2022masked}
Bowen Cheng, Ishan Misra, Alexander~G Schwing, Alexander Kirillov, and Rohit Girdhar.
\newblock Masked-attention mask transformer for universal image segmentation.
\newblock In \emph{Proceedings of the IEEE/CVF conference on computer vision and pattern recognition}, pp.\  1290--1299, 2022.

\bibitem[Hu et~al.(2025)Hu, Eidex, Wang, Safari, Li, and Yang]{hu2025benchmarking}
Mingzhe Hu, Zach Eidex, Shansong Wang, Mojtaba Safari, Qiang Li, and Xiaofeng Yang.
\newblock Benchmarking gpt-5 for zero-shot multimodal medical reasoning in radiology and radiation oncology.
\newblock \emph{arXiv preprint arXiv:2508.13192}, 2025.

\bibitem[Jin et~al.(2022)Jin, Huang, Zhou, Li, Yan, Sun, Zhang, Wang, and Ye]{jin2022fives}
Kai Jin, Xingru Huang, Jingxin Zhou, Yunxiang Li, Yan Yan, Yibao Sun, Qianni Zhang, Yaqi Wang, and Juan Ye.
\newblock {FIVES}: A fundus image dataset for artificial intelligence based vessel segmentation.
\newblock \emph{Scientific Data}, 9\penalty0 (1):\penalty0 475, 2022.

\bibitem[Kirillov et~al.(2023)Kirillov, Mintun, Ravi, Mao, Rolland, Gustafson, Xiao, Whitehead, Berg, Lo, et~al.]{kirillov2023segment}
Alexander Kirillov, Eric Mintun, Nikhila Ravi, Hanzi Mao, Chloe Rolland, Laura Gustafson, Tete Xiao, Spencer Whitehead, Alexander~C Berg, Wan-Yen Lo, et~al.
\newblock Segment anything.
\newblock In \emph{Proceedings of the IEEE/CVF international conference on computer vision}, pp.\  4015--4026, 2023.

\bibitem[Li et~al.(2023)Li, Wong, Zhang, Usuyama, Liu, Yang, Naumann, Poon, and Gao]{li2023llava}
Chunyuan Li, Cliff Wong, Sheng Zhang, Naoto Usuyama, Haotian Liu, Jianwei Yang, Tristan Naumann, Hoifung Poon, and Jianfeng Gao.
\newblock {LLaVA-Med}: Training a large language and vision assistant for biomedicine in one day.
\newblock In \emph{Advances in Neural Information Processing Systems}, volume~36, 2023.

\bibitem[Ma et~al.(2024)Ma, He, Li, Han, You, and Wang]{ma2023segment}
Jun Ma, Yuting He, Feifei Li, Lin Han, Chenyu You, and Bo~Wang.
\newblock Segment anything in medical images.
\newblock \emph{Nature Communications}, 15\penalty0 (1):\penalty0 654, 2024.

\bibitem[Mayo et~al.(2023)Mayo, Cummings, Lin, Gutfreund, Katz, and Barbu]{mayo2023hard}
David Mayo, Jesse Cummings, Xinyu Lin, Dan Gutfreund, Boris Katz, and Andrei Barbu.
\newblock How hard are computer vision datasets? calibrating dataset difficulty to viewing time.
\newblock \emph{Advances in Neural Information Processing Systems}, 36:\penalty0 11008--11036, 2023.

\bibitem[Moglia et~al.(2025)Moglia, Leccardi, Cavicchioli, Maccarini, Marcon, Mainardi, and Cerveri]{moglia2025generalist}
Andrea Moglia, Matteo Leccardi, Matteo Cavicchioli, Alice Maccarini, Marco Marcon, Luca Mainardi, and Pietro Cerveri.
\newblock Generalist models in medical image segmentation: A survey and performance comparison with task-specific approaches.
\newblock \emph{arXiv preprint arXiv:2506.10825}, 2025.

\bibitem[Ronneberger et~al.(2015)Ronneberger, Fischer, and Brox]{ronneberger2015u}
Olaf Ronneberger, Philipp Fischer, and Thomas Brox.
\newblock U-net: Convolutional networks for biomedical image segmentation.
\newblock In \emph{International conference on medical image computing and computer-assisted intervention}, pp.\  234--241. Springer, Cham, 2015.

\bibitem[Tajbakhsh et~al.(2015)Tajbakhsh, Gurudu, and Liang]{tajbakhsh2015automated}
Nima Tajbakhsh, Suryakanth~R Gurudu, and Jianming Liang.
\newblock Automated polyp detection in colonoscopy videos using shape and context information.
\newblock \emph{IEEE transactions on medical imaging}, 35\penalty0 (2):\penalty0 630--644, 2015.

\bibitem[Team et~al.(2023)Team, Anil, Borgeaud, Alayrac, Yu, Soricut, Schalkwyk, Dai, Hauth, Millican, et~al.]{team2023gemini}
Gemini Team, Rohan Anil, Sebastian Borgeaud, Jean-Baptiste Alayrac, Jiahui Yu, Radu Soricut, Johan Schalkwyk, Andrew~M Dai, Anja Hauth, Katie Millican, et~al.
\newblock Gemini: a family of highly capable multimodal models.
\newblock \emph{arXiv preprint arXiv:2312.11805}, 2023.

\bibitem[Tu et~al.(2024)Tu, Azizi, Driess, Schaekermann, Amin, Chang, Carroll, Lau, Tanno, Ktena, et~al.]{tu2024medpalm}
Tao Tu, Shekoofeh Azizi, Danny Driess, Mike Schaekermann, Mohamed Amin, Pi-Chuan Chang, Andrew Carroll, Charles Lau, Ryutaro Tanno, Ira Ktena, et~al.
\newblock Towards generalist biomedical ai.
\newblock \emph{Nejm Ai}, 1\penalty0 (3):\penalty0 AIoa2300138, 2024.

\bibitem[Vallez et~al.(2025)Vallez, Bueno, Deniz, Rienda, and Pastor]{vallez2025bus_uclm}
Noelia Vallez, Gloria Bueno, Oscar Deniz, Miguel~Angel Rienda, and Carlos Pastor.
\newblock Bus-uclm: Breast ultrasound lesion segmentation dataset.
\newblock \emph{Scientific Data}, 12\penalty0 (1):\penalty0 242, February 2025.
\newblock \doi{10.1038/s41597-025-04562-3}.

\bibitem[Wu et~al.(2023)Wu, Lei, Zheng, Zhao, Lin, Zhang, Zhou, Zhao, Zhang, Wang, et~al.]{wu2023can}
Chaoyi Wu, Jiayu Lei, Qiaoyu Zheng, Weike Zhao, Weixiong Lin, Xiaoman Zhang, Xiao Zhou, Ziheng Zhao, Ya~Zhang, Yanfeng Wang, et~al.
\newblock Can gpt-4v (ision) serve medical applications? case studies on gpt-4v for multimodal medical diagnosis.
\newblock \emph{arXiv preprint arXiv:2310.09909}, 2023.

\bibitem[Zenk et~al.(2025)Zenk, Zimmerer, Isensee, Traub, Norajitra, J{\"a}ger, and Maier-Hein]{zenk2024comparative}
Maximilian Zenk, David Zimmerer, Fabian Isensee, Jeremias Traub, Tobias Norajitra, Paul~F J{\"a}ger, and Klaus Maier-Hein.
\newblock Comparative benchmarking of failure detection methods in medical image segmentation: unveiling the role of confidence aggregation.
\newblock \emph{Medical image analysis}, 101:\penalty0 103392, 2025.

\bibitem[Zhang et~al.(2022)Zhang, Fu, Zheng, Zhang, Zhao, and Sham]{zhang2022hsnet}
Wenchao Zhang, Chong Fu, Yu~Zheng, Fangyuan Zhang, Yanli Zhao, and Chiu-Wing Sham.
\newblock Hsnet: A hybrid semantic network for polyp segmentation.
\newblock \emph{Computers in biology and medicine}, 150:\penalty0 106173, 2022.

\bibitem[Zhang et~al.(2024)Zhang, Pan, Zhong, Dong, Xie, Liu, Jiang, Wu, Liu, Zhao, et~al.]{zhang2024potential}
Yutong Zhang, Yi~Pan, Tianyang Zhong, Peixin Dong, Kangni Xie, Yuxiao Liu, Hanqi Jiang, Zihao Wu, Zhengliang Liu, Wei Zhao, et~al.
\newblock Potential of multimodal large language models for data mining of medical images and free-text reports.
\newblock \emph{Meta-Radiology}, 2\penalty0 (4):\penalty0 100103, 2024.

\end{thebibliography}
\bibliographystyle{tmlr}

\end{document}